\documentclass[twocolumn,a4paper]{article} 
\usepackage{multicol} 
\newcommand{\ea}{{\it et al.\ }} 
\begin{document} 
\twocolumn[\hsize\textwidth\columnwidth\hsize\csname @twocolumnfalse\endcsname 
\title{\bf Spin-polarized muons in condensed matter physics} 
\author{{\scshape S. J. Blundell}\\ \\ 
Oxford University Department of Physics,\\ 
Clarendon Laboratory, Parks Road, Oxford OX1 3PU} 
\maketitle 
\begin{abstract} 
{\it 
A positive muon is a spin-$1/2$ particle.  Beams of muons 
with all their spins polarized can be prepared and subsequently 
implanted in various types of condensed matter.  The 
subsequent precession and relaxation of their spins 
can then be used to investigate a variety of 
static and dynamic effects in a sample and 
hence to deduce properties concerning magnetism, 
superconductivity and molecular dynamics. 
Though strictly a lepton, and behaving essentially 
like a heavy electron, it is convenient to think of 
a muon as a light proton, and it is often found with 
a captured electron in a hydrogen-like atom 
known as muonium. 
This article outlines the principles of various 
experimental techniques which involve implanted muons 
and describes some 
recent applications.  The use of muons 
in condensed matter physics has shed new light on 
subjects as diverse as 
passivation in semiconductors, frustrated spin systems, 
vortex lattice melting, and quantum diffusion of 
light particles. 
}
\end{abstract} 
\begin{center}
Originally published in {\it Contemporary Physics} {\bf 40},  175,  (1999)
\end{center}
\vskip 0.5cm 

] 
 
\section{Introduction} 
\begin{figure} 
\vspace{6cm} 
\includegraphics{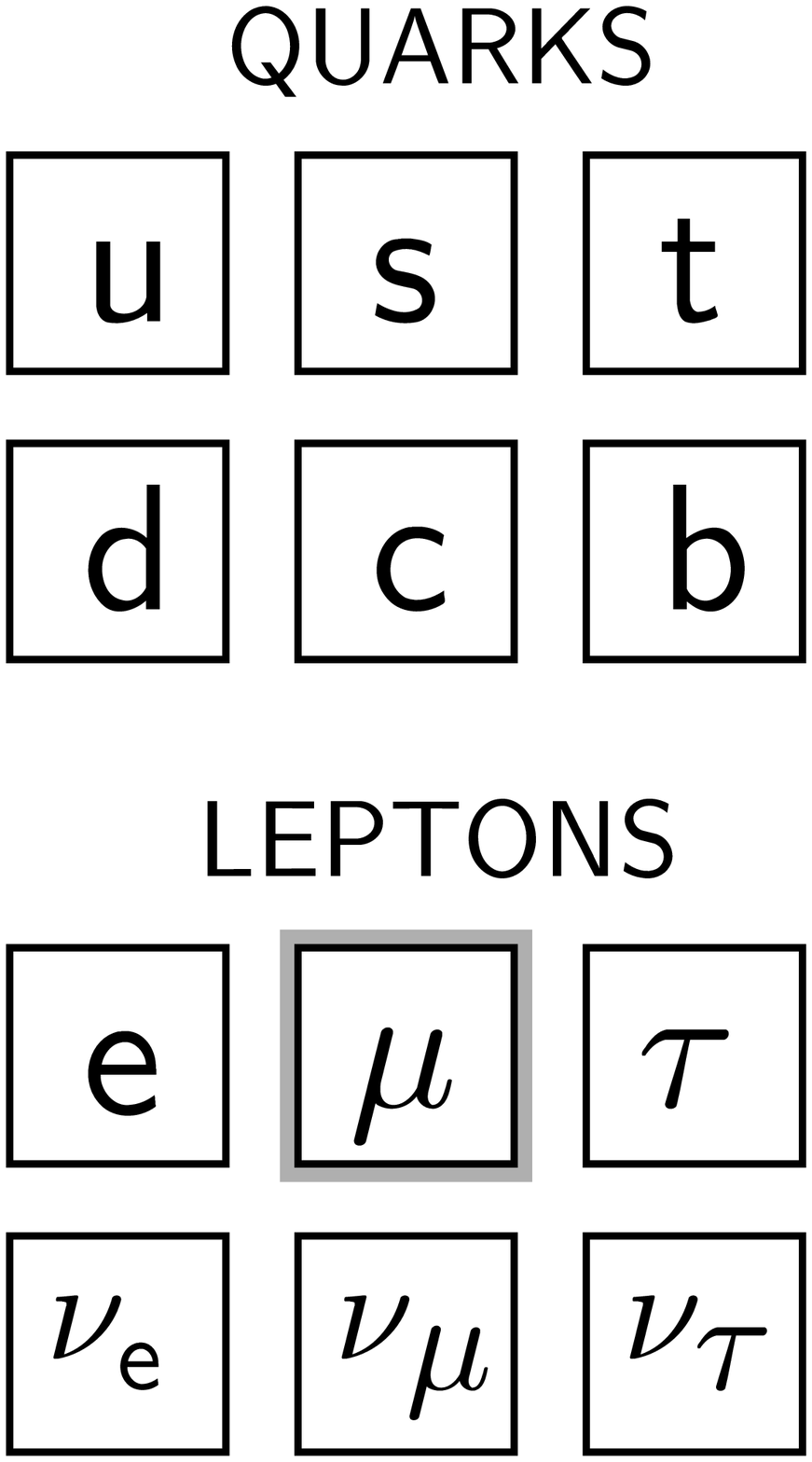} 
\caption{Quarks ($u$=up, $d$=down, $s$=strange, $c$=charm, 
$t$=top, $b$=bottom) and leptons ($e$=electron, $\mu$=muon, 
$\tau$=tau, and their associated neutrinos $\nu_e$, $\nu_\mu$, $\nu_\tau$. 
Everyday matter is composed of particles from the first column 
in this table.} 
\label{partic} 
\end{figure}

Condensed matter physics is concerned with the properties 
of solids, liquids and various intermediate states 
of matter such as colloids (Chaikin and Lubensky 1995). 
From a fundamental viewpoint, 
all matter is  made up of two main types 
of constituent particles, quarks and leptons (see Figure~\ref{partic}). 
The universe is composed of only these particles 
in addition to the gauge bosons (such as photons) which mediate 
forces between particles. 
Protons and 
neutrons are each made up of three quarks while electrons 
are thought to be indivisible and are members of 
a class of particle called leptons. 
Everyday matter, made up of 
protons, neutrons and electrons, is the main concern 
of condensed matter physics and therefore the 
subject arises from particles in the 
first column (or `generation') of the standard model (Figure~\ref{partic}). 
However, more exotic particles do exist which are either 
more unusual leptons or 
combinations of more exotic quarks; they are all 
short-lived and have never been found to be relevant in condensed 
matter physics with one single exception: the muon. 
 
The muon {\it is} found in nature; it is the dominant constituent 
of the cosmic rays arriving at sea-level and a few will be 
hitting you each minute as you read this article.  The fact that 
so many muons make it through the atmosphere is a well-known 
consequence of time-dilation effects in special relativity 
(Rossi and Hall 1941, Bailey \ea 1977).  However, at a number of locations 
in the world, 
intense beams of muons are prepared artificially for 
research in condensed matter physics.  In this article I 
will try and explain why this is done and what can be 
learned by firing this rather exotic `second-generation' 
particle into various types of matter. 
I will describe how implanted muon studies have 
shed new light on problems in condensed matter physics, 
including those concerning semiconductors, magnetism, 
superconductors, and quantum diffusion. 
A key difference between this technique and those involving 
neutrons and X-rays is that scattering is {\it not} involved. 
Neutron diffraction methods use the change in energy and/or 
momentum of a scattered neutron to infer details about a 
particular sample.  In contrast muons are {\it implanted} 
into a sample and reside there for the rest of their lives. 
The muons 
themselves never emerge again.  It is the positrons into which they 
decay that are released from the sample 
and yield information 
about the muons from which they came. 
The experimenter plays the r\^ole of a coroner at an inquest; 
the evidence of the decay particles given up at death allows 
one to infer the nature of the muon's life.

\section{The discovery of the muon} 
Although the discovery of the muon is usually considered 
as a single event which  is credited 
to Neddermeyer and Anderson because of their work in 1936, 
the story is actually more complicated and one might 
say that the muon slowly ``emerged'' over almost a half-century. 
Following the discovery of radioactivity in 1899, it was 
observed that electrometers, instruments capable of measuring 
the ionization produced by radioactivity, 
discharged 
even when there was no obvious radioactive source. 
A Jesuit priest called Thomas Wulf 
noticed in 1910 that this effect was more 
pronounced at the top of the Eiffel tower than at the bottom 
(Wulf 1910, see also Xu and Brown 1987).  That 
the rate of discharge of an electrometer was an increasing function 
of altitude was demonstrated beyond doubt by Victor Hess in 
an intrepid series of ballooning experiments in the period 1911--12 
(Hess 1912). 
He showed that electrometers discharged between three and five 
times faster at 
an altitude of 5000~m than they did at sea level. 
Such experiments were later extended to very high altitudes by the 
use of remote-controlled electrometers (in which the experimenter 
did not actually accompany their experiment in the balloon); this 
advance was achieved by Robert Millikan who was initially  sceptical 
about Hess' work but became rapidly convinced of its importance 
(Xu and Brown 1987). 
Millikan's name for the effect, {\it cosmic rays}, gained 
universal acceptance, but it was Hess who was awarded the Nobel prize for 
the discovery in 1936.  The major constituent of cosmic rays 
at ground level 
turned out to be the muon, but the true identity of the 
muon remained hidden for some time. 
 
Millikan's group at CalTech began to perform experiments on cosmic 
rays at ground level by bending the rays in magnetic and electric 
fields.  This work led to the discovery of the positron 
by Carl Anderson in 1932 (Anderson shared the Nobel 
prize in 1936 with Hess) and then the muon (initially called 
the {\it mesotron}, and later the {\it mu-meson}) was identified 
in cosmic rays by 
Seth Neddermeyer and Carl Anderson in 1936.  They measured the 
mass of the muon, 
which turned out to be roughly one-ninth of the proton mass, and 
agreed extremely well with the predictions of 
the Japanese physicist 
Hideo Yukawa for a particle to mediate the strong force and 
hold the nucleons in the nucleus together. 
This result can be understood 
by using Heisenberg's uncertainty principle $\Delta E\Delta t \sim \hbar$ 
for this virtual particle. 
$\Delta E$ is the 
the rest mass energy of Yukawa's particle, the energy 
which has to be ``borrowed'', and $\Delta t$ is the 
time for it to cross the nucleus, the time for which it has to be 
``borrowed''; by assuming a speed $\sim c$ for the particle 
one arrives at a mass $\sim$100~MeV, the mass found for the muon. 
The muon was therefore 
assumed to be the particle which holds the nucleus together. 
It took some technically difficult experiments to show that this 
assumption was in error and that the startling 
agreement with Yukawa's predictions was misleading. 
Marcello Conversi, Ettore Pancini and Oreste Piccioni, working 
in a basement in Rome during and after the Second World War, 
tried to implant cosmic ray muons 
in matter and then measure their lifetime (Conversi \ea 1945, 1947). 
They found that positive muons implanted 
in anything always live on average for 2.2~$\mu$s (the lifetime 
in vacuum); the lifetime 
of negative 
muons on the other hand depends on the atomic number $Z$ of the 
material into which it is implanted; 
they measured 2.2~$\mu$s for carbon but found 
0.07~$\mu$s for lead.  They reasoned that 
if muons really mediated the strong force 
they should be gobbled up much more quickly in all materials. 
It appeared that muons interacted only rather weakly with matter and 
the only effect that was observed occurred for negative 
muons and turned out to be $\mu^-$ capture: 
$$ \mu^- + p \to n + \nu_\mu.$$ 
A negative muon is attracted by atomic nuclei and since its 
mass is much greater than that of an electron, it readily 
displaces an electron from an atom and rapidly drops down 
to the $1s$ state.  From there it either decays to an electron 
or undergoes capture as above. 
The propensity to undergo capture depends on the atomic number 
$Z$ (approximately as the fourth power of $Z$) 
through  the Bohr radius of the $\mu^-$ orbit. 
 
Yukawa's particle turned out to be the pion, discovered in 
cosmic rays by Powell's group in 1947 (Lattes \ea 1947), which decays into a muon 
after only 0.026~$\mu$s, making it harder to spot in cosmic rays. 
Thus the muon was not a meson (a term now reserved for 
quark--anti-quark pairs like the pion) but a lepton, a heavy electron 
with no internal structure. 
 
\begin{figure} 
\vspace{3.75cm} 
\includegraphics{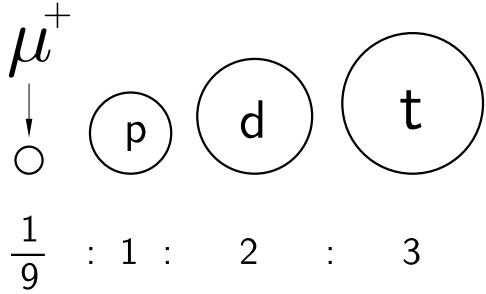} 
\caption{ 
The muon can be considered to be the lightest member of 
the family of particles which include the proton, deuteron 
and triton.  The approximate ratio of their masses are as shown. 
The muon thus extends the mass range of isotope substitution
available to the experimenter.
} 
\label{isotopes} 
\end{figure} 
 
For experiments in 
condensed matter physics 
it is mainly the {\it positive} muon which is used. 
The negative muon $\mu^-$, 
which implants close to an atomic nucleus, is generally 
much less sensitive to the phenomena of interest  to 
condensed matter physicists (magnetism, 
superconductivity, etc) than the site of the implanted 
positive muon $\mu^+$ which 
sits well away from nuclei in regions of large electron 
density.  Condensed matter physics is essentially the 
physics of electrons, rather than nuclei, so that the best 
place you can put your test particle is in the electron cloud. 
In fact even though the muon is a 
lepton (see Figure~\ref{partic}) and therefore essentially a heavy 
electron, for our purpose it is more useful to 
consider it as a light proton (see Figure~\ref{isotopes}). 
 
\begin{figure} 
\vspace{4cm} 
\includegraphics{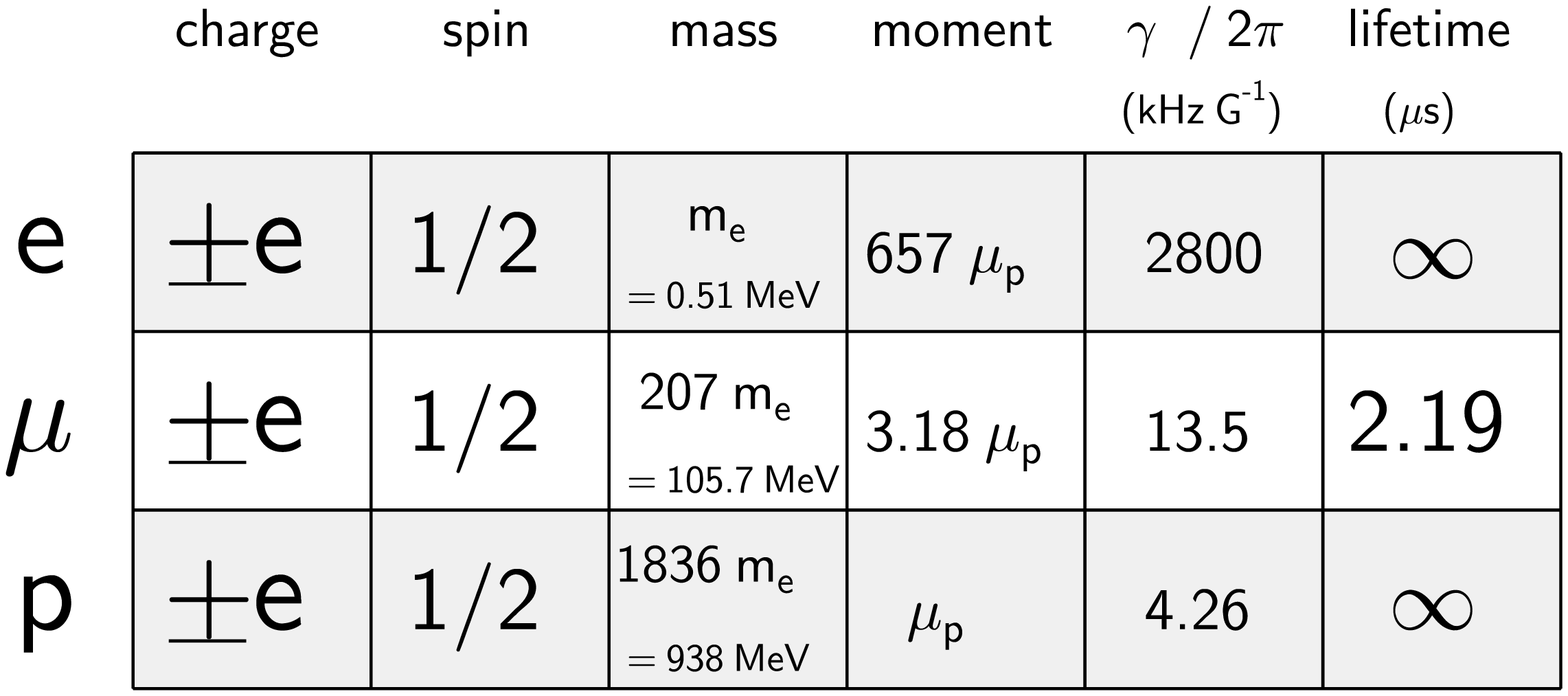} 
\caption{Properties of the electron, muon and proton.} 
\label{props} 
\end{figure}

Some properties of the electron, muon and proton are tabulated in 
Figure~\ref{props}.  The mass of the muon is intermediate between 
that of the electron and the proton, and thus so are its 
magnetic moment and gyromagnetic ratio. 
The latter is the constant of proportionality between angular momentum 
and magnetic moment. 
 
\section{Muon production, decay and implantation} 
Cosmic rays provide a major source of muons 
[roughly one muon arrives vertically on each square 
centimetre of the earth's surface every minute (Caso \ea 1998)] 
but, since the 1950s, experiments on muons have 
needed higher intensities and have 
required the development of accelerators. 
High energy proton beams (produced using synchrotrons 
or cyclotrons) are fired into a target (usually 
graphite) to produce pions via 
$$ p+p \to \pi^+ + p + n, $$ 
and the pions decay into muons: 
$$ \pi^+ \to \mu^+ + \nu_\mu, $$ 
where $\nu_\mu$ is a muon-neutrino. 
The pion decay is a two-body decay and is therefore particularly 
simple.  For example, consider the pions which are produced at rest 
in the laboratory frame.  To conserve momentum, the muon 
and the neutrino must have equal and opposite momentum. 
The pion has zero spin so the muon spin must be opposite to 
the neutrino spin.  One useful property of the neutrino 
is that its spin is aligned antiparallel with its momentum (it 
has negative helicity), 
and this implies that the muon-spin is similarly aligned. 
Thus by selecting pions which stop in the target (and which 
are therefore at rest when they decay) one has a means of 
producing a beam of 100\% spin-polarized muons. 
This is the method most commonly used for producing muon 
beams for condensed matter physics research, though other 
configurations are in use (Brewer 1994).

\begin{figure} 
\vspace{11.5cm} 
\includegraphics{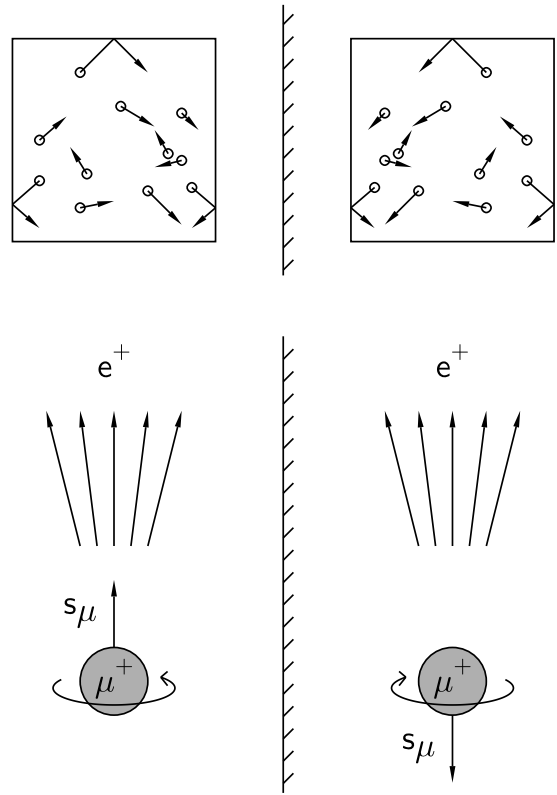} 
\caption{ 
Top: molecules in a box collide with the walls.  These collisions 
do not violate parity so that both the process shown on the 
left, and its mirror image on the right, could be observed in nature. 
Bottom: the same is not true for the process of muon decay. 
The direction of the 
muon-spin is reversed in the mirror so that the 
positrons are emitted predominantly in a direction 
opposite to that of the muon-spin. 
The violation of parity means that in our universe 
only the process on the left-hand side of the diagram 
is ever observed. 
} 
\label{parity} 
\end{figure}

The muons are stopped in the specimen of interest 
and decay after a time $t$ with probability proportional to 
$e^{-t/\tau_\mu}$ where $\tau_\mu=2.2$~$\mu$s is 
the lifetime of the muon. 
The muon decay is a three body process 
$$ \mu^+ \to e^+ + \nu_e + \bar{\nu}_\mu $$ 
and so the energy of the positron $e^+$ (which is the 
only particle produced in this reaction that 
we have a sensible hope of reliably detecting) 
may vary depending on how  momentum is distributed 
between the three particles (subject to the constraint 
that the total vector 
momentum will sum to zero, the initial momentum of the 
stopped muon). 
The decay involves the weak interaction and thus 
has the unusual feature of not conserving parity (Garwin \ea 1957). 
This phenomenon (which also lies behind the negative helicity 
of the neutrino) leads to a propensity for the emitted positron 
to emerge predominantly along the direction of the muon-spin 
when it decayed.  This 
can be understood by considering the 
mirror image of the muon decay process (Figure~\ref{parity}). 
 
\begin{figure} 
\vspace{5cm} 
\includegraphics{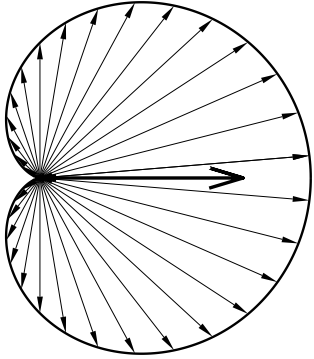} 
\caption{The angular distribution of emitted positrons with 
respect to the initial muon-spin direction.  The figure shows 
the expected distribution for the most energetically emitted 
positrons.} 
\label{asy} 
\end{figure}

The angular distribution of emitted positrons is shown in 
Figure~\ref{asy} for the case of the most energetically emitted 
positrons.  In fact positrons over a range of energies 
are emitted so that the net effect is something not quite 
as pronounced, but the effect nevertheless allows one to follow the 
polarization of an 
ensemble of precessing muons with arbitrary accuracy, providing 
one is willing to take data for long enough. 
 
Muons are implanted into the sample with an energy which is 
at least 4~MeV.  They lose energy very quickly (in 0.1--1~ns) 
to a few keV 
by ionization of atoms and scattering with electrons. 
Then the muon begins to undergo a series of 
successive electron capture and loss reactions which 
reduce the energy to a few hundred eV in about a picosecond. 
If muonium is ultimately formed then electron capture 
ultimately wins and the last few eV are shed by 
inelastic collisions between the muonium atom and 
the host atoms.  All of these effects are very fast 
so that the muon (or muonium) is thermalized very rapidly. 
Moreover the effects are all Coulombic in origin and do 
not interact with the muon-spin so that 
the muon is thermalized in matter without appreciable 
depolarization. 
This is a crucial feature for muon-spin rotation experiments. 
One may be concerned  that the muon may only measure a region of 
sample which has been subjected to radiation damage by 
the energetic incoming muon.  This does not appear to be 
a problem since there is a threshold energy for 
vacancy production, which means that only the initial 
part of the muon path suffers much damage.  Beyond this 
point of damage the muon still has sufficient energy to 
propagate through the sample a further distance  thought to be 
about 1~$\mu$m, leaving it well away from any induced 
vacancies (Chappert 1984). 
 
\section{Spin precession and relaxation} 
 
\begin{figure} 
\vspace{8cm} 
\includegraphics{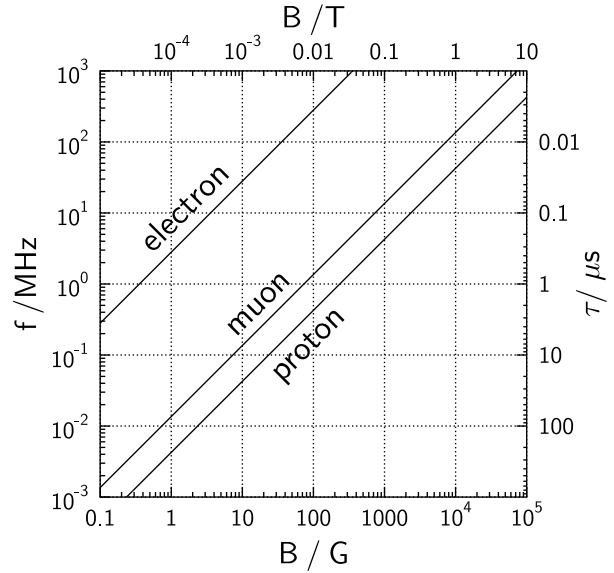} 
\caption{The Larmor precession frequency $f$ in MHz (and the 
corresponding period $\tau=1/f$) for the electron, muon 
and proton as a function of applied magnetic field $B$.} 
\label{precess} 
\end{figure}

In a magnetic field $B$ the muon-spin precesses with 
angular frequency $\omega_\mu$ given by $\omega_\mu = \gamma_\mu B$ 
where $\gamma_\mu = ge/2m_\mu$ is the gyromagnetic 
ratio for the muon.  This is known as Larmor precession. 
The field-dependent precession frequencies for the muon, electron and proton 
are  shown in Figure~\ref{precess}.  The highest frequencies 
are associated with the lightest particle, the electron, and 
the lowest with the proton.  For usual laboratory magnetic fields this 
explains why 
ESR (electron-spin resonance) is typically performed at microwave 
frequencies while NMR (nuclear magnetic resonance) uses radio-frequencies. 
In both of these techniques resonance occurs when the precession frequency 
matches the resonance frequency. 
Muon-spin rotation ($\mu$SR) is associated with frequencies 
intermediate between NMR and ESR  but unlike those resonance 
techniques, no electromagnetic field is necessary 
since the precessing muon can be followed directly. 
[Muon-spin {\it resonance} experiments 
can however be performed (Brewer 1994) but a discussion is 
outside the scope of this article.] 
 
\begin{figure} 
\vspace{8cm} 
\includegraphics{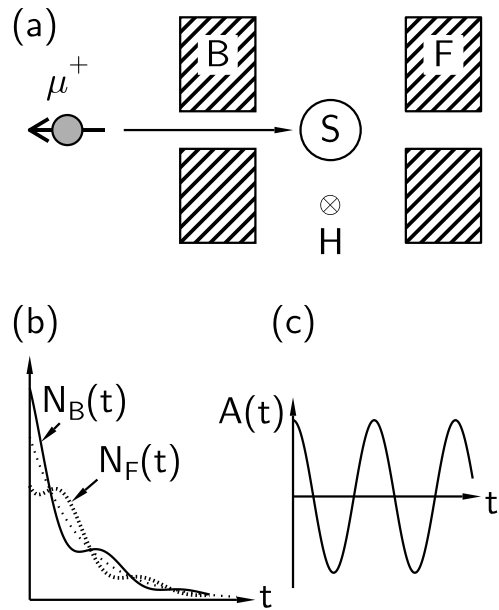} 
\caption{Schematic illustration of a $\mu$SR experiment. 
(a) A spin-polarized beam of muons is implanted in a sample $S$. 
Following decay, positrons are detected in either 
a forward detector $F$ or a backward detector $B$. If a transverse 
magnetic field $H$ is applied to the sample as shown then the 
muons will precess. (b) The number of positrons detected in the 
forward and backward detectors. (c) The asymmetry function. 
} 
\label{expt} 
\end{figure} 
 
A schematic diagram of the experiment is shown in 
Figure~\ref{expt}(a).  A muon, with its polarization aligned 
antiparallel to its momentum, is implanted in a sample. 
(It is antiparallel because of the way that it was formed, see above, 
so the muon enters the sample with its spin pointing along the 
direction from which it came.) 
If the muon is unlucky enough to decay immediately, then it will 
not have time to precess and a positron will be emitted 
preferentially into 
the backward detector.  If it lives a little longer it will have time 
to precess so that if it lives for half a revolution the resultant 
positron will be preferentially emitted into the forward detector. 
Thus the positron beam from an ensemble of precessing muons can 
be likened to the beam of light from a lighthouse. 
 
The time evolution of the number of positrons detected in 
the forward and backward detector is described by the functions 
$N_F(t)$ and $N_B(t)$ respectively and these are shown in 
Figure~\ref{expt}(b). 
Because the muon decay is a radioactive process 
these two terms sum to an exponential decay.  Thus the 
time evolution of the 
muon polarization can be obtained by examining the normalized difference 
of these two functions via the asymmetry function $A(t)$, given by 
\begin{equation} 
A(t)={N_B(t)-N_F(t) \over N_B(t)+N_F(t)}, 
\end{equation} 
and is shown in Figure~\ref{expt}(c).

This experimentally obtained asymmetry function has a 
calculable  maximum value, $A_{\rm max}$, 
for a particular experimental configuration which 
depends on the initial beam polarization (usually very close to 1), 
the intrinsic asymmetry of the weak decay, and the 
efficiency of the detectors for positrons of different energies, 
and usually turns out to be around $A_{\rm max}\sim$0.25.  The function 
can be normalized to 1,  in which case it expresses 
the spin autocorrelation function of the muon, $G(t)=A(t)/A_{\rm max}$, 
which represents the time-dependent spin polarization of the muon.

\begin{figure} 
\vspace{5.7cm} 
\includegraphics{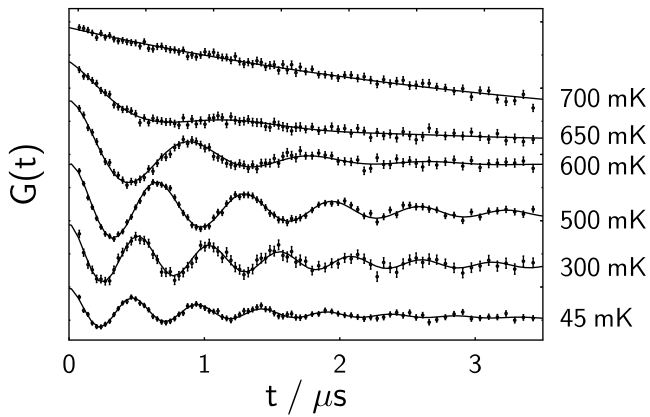} 
\caption{Zero-field muon spin rotation frequency 
in the organic ferromagnet {\it p}-NPNN (Blundell \ea 1995). 
} 
\label{npnn} 
\end{figure}

A magnetic field does not need to be applied for the muons 
to precess if the sample has its own magnetic field. 
Figure~\ref{npnn} shows muon-spin rotation data for muons 
implanted into an organic ferromagnet (Blundell \ea 1995). 
This material magnetically orders only at very low temperatures 
($T_{\rm C}= 0.67$~K) 
so that experiments must be carried out in a dilution 
refrigerator.  This presents no problems for the muon which 
passes through the windows of the cryostat and implants 
in the sample, revealing the internal magnetisation. 
As the sample is warmed, the frequency of oscillations decreases 
as the internal field decreases until it is above the Curie 
temperature and no oscillations can be observed, only a 
weak spin relaxation arising from spin fluctuations. 
 
This is an example of an experiment with no applied magnetic 
field (in fact a small magnetic field was applied to compensate 
the effect of the earth's field).  Very often magnetic fields 
are applied to the sample either perpendicular or parallel 
to the initial muon-spin direction.  The perpendicular (or transverse) 
case causes the muon to precess in the applied magnetic field 
and any dephasing in the observed oscillations is evidence for 
either an inhomogeneous internal field distribution or 
spin-spin ($T_2$ in the language of NMR) relaxation.  The parallel 
(or longitudinal) 
case will be described in more detail later and does not 
lead to spin precession, but spin relaxation.  This can 
be due to inhomogeneous field distributions or 
spin-lattice ($T_1$) relaxation processes. 
 
\begin{figure} 
\vspace{3.5cm} 
\includegraphics{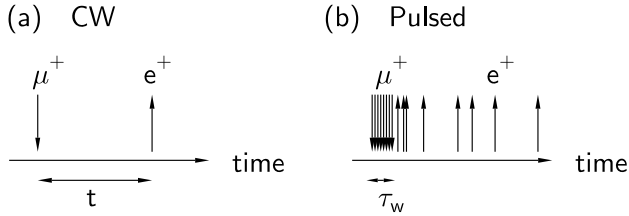} 
\caption{Schematic illustration of the two types of muon beam 
(a) continuous wave (CW) and (b) pulsed.} 
\label{muontime} 
\end{figure}

Muon experiments can be performed in two different ways depending 
on the time structure of the muon beam.  If the muon beam is 
continuous (CW, or continuous wave 
see Figure~\ref{muontime}(a)), then muons arrive at the 
sample intermittently.  When the muon enters the experiment it 
must itself be detected to start a clock.  When the positron 
is detected in either the forward or backward detectors, the clock is stopped. 
If a second muon arrives before the first one has decayed then 
one has no way of knowing whether a subsequently emitted positron 
came from the first or second muon, so this event must be 
disregarded.  Sophisticated high-speed electronics and a 
low incident muon arrival rate are needed.  Alternatively 
one can use an electrostatic deflector triggered by the 
detectors to ensure no muons enter the experiment until 
the current implanted muon decays. 
 
These complications 
are circumvented with a pulsed muon beam (Figure~\ref{muontime}(b)). 
In this case a large number of muons 
arrive in a very intense pulse so there is no need to detect 
when each muon arrives.  The detection of positrons is then 
made and each event is timed with respect to the arrival of the 
pulse. 
A typical dataset contains several million detected positrons so 
that 
an appreciable number of muons (the fraction is given by 
$e^{-20/2.2}\sim$0.01\%) live for 20~$\mu$s or longer. 
Long-lived muons are difficult to measure with CW beams; the 
the arrival of the next muon tends to interrupt the first muon which has 
outstayed its welcome!  Nevertheless the long-lived muons 
can be accurately detected at a pulsed source. 
Unfortunately this method also suffers from a drawback 
which is that the muon pulse has a finite width, $\tau_w$, 
which results 
in a slight ambiguity in all of the timing measurements 
and leads to an upper limit on precession frequencies which 
can be measured. 
By the standard of other unstable elementary particles, the 
muon is comparatively long-lived with a lifetime of $\tau=$2.2~$\mu$s. 
CW muon beams are operated at the Paul Scherrer Institute in 
Switzerland and at TRIUMF in Canada.  Pulsed muon beams are 
used at KEK in Japan and at ISIS (the spallation source at 
the Rutherford Appleton Laboratory) in the UK. 
 
\section{Muonium} 
\label{muonium} 
\begin{figure} 
\vspace{6.2cm} 
\includegraphics{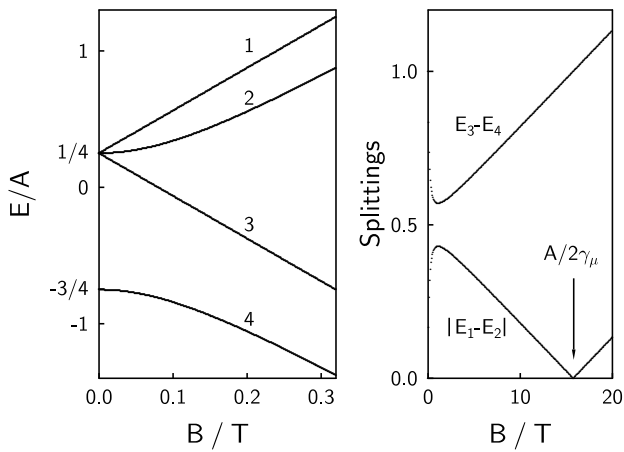} 
\caption{ 
The Breit-Rabi diagram for isotropic muonium with a 
hyperfine constant $A=4463$~MHz appropriate for 
free (vacuum) muonium.  Also shown are the splittings 
between pairs of levels over a larger range of magnetic 
field.  Levels 1 and 2 cross at a field of $\sim$16.5~T 
for free muonium. 
} 
\label{breit} 
\end{figure}

Depending on its chemical environment, the muon can thermalize 
and pick up an electron and form a neutral atomic 
state called muonium (abbreviated Mu$=\mu^+e^-$) which is 
an analogue of atomic hydrogen.  In muonium the electronic spin 
and the muon-spin are coupled by a hyperfine interaction 
which we will initially assume is isotropic.  This leads 
to two energy levels, a lower triplet state and a higher singlet state. 
In a magnetic field the triplet levels split and the energy levels 
move as shown in the Breit-Rabi diagram in Figure~\ref{breit}. 
 
\begin{figure} 
\vspace{5cm} 
\includegraphics{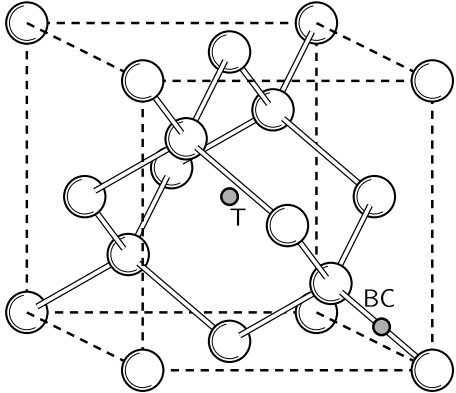} 
\caption{ 
The crystal structure of silicon, showing the 
possible muonium sites (T=tetrahedral site, BC=bond-centre site). 
} 
\label{silicon} 
\end{figure}

Isolated hydrogen or muonium is one of the simplest defects 
in a semiconductor.  The electronic structures of analogous 
hydrogen and muonium centres are expected to be identical, 
apart from vibrational effects reflecting their difference 
in mass. 
If muons are implanted into a semiconductor like silicon, neutral 
muonium tends to form at the tetrahedral sites (labelled $T$ 
in Figure~\ref{silicon}) and can diffuse rapidly between such sites. 
The electronic state is isotropic and in a transverse-field experiment 
the transitions between energy levels can be observed as 
precession frequencies $\nu_{ij}$ equal to the 
splitting between energy levels $(E_i-E_j)/h$ 
(as shown in Figure~\ref{breit}). 
The strength of the hyperfine interaction in semiconductors is usually close 
to half that of the vacuum value (the reduction is due to some admixture 
of the electron spin density with surrounding atoms).
However, a substantial fraction of neutral muonium 
is also found in a most unexpected place, wedged into the centre of a stretched 
Si--Si bond 
(the site is labelled $BC$ in Figure~\ref{silicon} for `bond-centre', 
see Patterson 1988 for a review of muonium states in semiconductors). 
This state is extremely immobile, and surprisingly turns out to be 
the thermodynamically more stable site.  Its hyperfine coupling is 
much lower than that of the tetrahedral state, typically 
less than 10\% of the vacuum value. Furthermore the coupling is 
very anisotropic, with axial symmetry 
about the $\langle 111\rangle$ crystal axis 
(i.e.\ along the Si--Si bonds), 
so that the energy levels behave in a different 
manner to that indicated in Figure~\ref{breit}. 
These states have rather interesting dynamics and can undergo charge 
and spin exchange processes, cycling rapidly between positive and 
negative charge states via interaction with conduction electrons (Chow 
\ea 1994). 
The characterisation of all these states is important because it 
is found that atomic hydrogen is present in most semiconductors 
and is able to passivate (i.e. deactivate) the dangling bonds 
in amorphous silicon allowing it to show semiconducting properties. 
Hydrogen is inevitably present in all semiconductors, often becoming 
incorporated during material production from hydride gases or during 
etching, but the low concentration makes direct spectroscopic 
studies very difficult. 
Using 
muonium (albeit in the ultra-dilute limit!) as an analogue for hydrogen 
has therefore been a promising method of obtaining a great deal of 
spectroscopic information concerning this problem. 
 
Muonium states can also be formed in many chemical systems 
and allow a unique form of radical spectroscopy (Roduner 1993). 
Muonium adds to unsaturated bonds to form muonated free radicals. 
For example, addition to benzene (C$_6$H$_6$) leads to 
the muonated cyclohexadienyl radical (C$_6$H$_6$Mu). 
The advantage here is that one can work with concentrations 
down to just one muonated radical at a time in an entire 
sample.  In contrast ESR detection needs $\sim 10^{12}$ 
radicals in a cavity, forbidding measurements at high 
temperatures where the radicals become mobile and 
terminate by combination.  The technique has been applied 
to radicals in various environments (Roduner 1993) including 
those absorbed on surfaces (Reid \ea 1990). 
 
\begin{figure} 
\vspace{5.3cm} 
\includegraphics{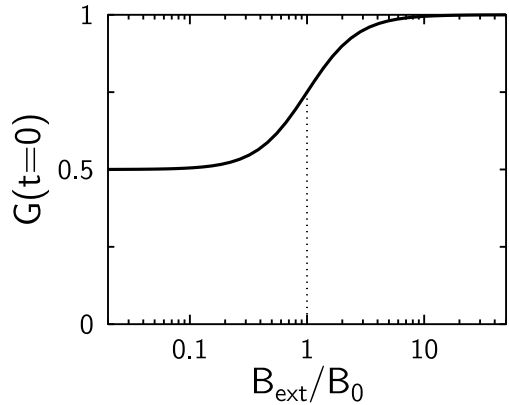} 
\caption{ 
Repolarization of the initial muon polarization. 
} 
\label{repol} 
\end{figure} 
 
In all cases the muonium can be studied by measuring precession 
signals in an applied magnetic field, or by using a technique 
known as repolarization (Figure~\ref{repol}). 
In this latter method a longitudinal magnetic field 
is applied to the sample, along the initial muon-spin direction, 
and as the strength of the magnetic field increases, the muon 
and electron spins are progressively decoupled from the 
hyperfine field.  For isotropic muonium, half of the initial 
polarization of implanted muons is lost because of the 
hyperfine coupling, but this is recovered in a sufficiently 
large applied field (Figure~\ref{repol}), allowing an estimate 
of the strength of the hyperfine field.  If the hyperfine 
coupling is anisotropic then this affects the form of 
the repolarization curve, allowing further information to 
be extracted (Pratt 1997). 
 
At very high fields, when the electron and muon spins are completely 
decoupled, the initial muon polarization is preserved.  However, 
at certain values of the magnetic field a level-crossing of 
energy levels may occur.  In fact, there is one at very high field 
in vacuum muonium (see Figure~\ref{breit}(b)). 
Interactions between the muon-electron system and nuclei in the 
host material cause the pure Zeeman states to mix near these 
level-crossings, thereby avoiding the crossing, 
and this can lead to a loss of 
polarization.  This provides a further technique to extract 
hyperfine coupling strengths and also measurements of quadrupole 
couplings 
(Cox 1987, Roduner 1993, Brewer 1994). 
It turns out that these avoided-level-crossing resonances 
are also a sensitive probe of dynamics and this technique has 
been applied to a number of studies of molecular motion (Roduner 1993). 
 
An interesting situation occurs in the case of C$_{60}$ 
in which muonium can implant inside the buckyball cage 
(this state is called endohedral muonium). 
The unpaired electron part of the muonium greatly enhances 
the sensitivity to scattering from conduction electrons 
so that this state is extremely useful for studying 
alkali-fulleride superconductors (MacFarlane \ea 1998). 
It is also possible to form a muonium radical by 
external addition, essentially muonium attacking the 
outside of a buckyball, breaking a double bond and 
ending up covalently bonded to a single saturated carbon 
atom.  This centre is very sensitive to the molecular 
dynamics of the local environment and has been used 
to extract the correlation time for molecular 
reorientation (Kiefl \ea 1992). 
 
In  metallic samples the muon's positive charge is screened by conduction 
electrons which form a cloud around the muon, of size 
given by a Bohr radius.  Thus $\mu^+$, rather than muonium, is the 
appropriate particle to consider in a metal. 
(The endohedral muonium found in alkali fulleride superconductors 
is the only known example of a muonium state in a metal.) 
In insulators and semiconductors  screening cannot take 
place so that the muon is often observed in these systems either 
as muonium or is found to be chemically bound to 
one of the constituents, particularly to oxygen if it is present. 
Isotropic muonium states are found in many semiconducting 
and insulating systems.  The value of the hyperfine coupling 
strength is 
close to that for vacuum (free) muonium if the band gap is large. 
For materials with smaller band gap the hyperfine coupling 
is lower reflecting the greater delocalization of the electron 
spin density on to neighbouring atoms (Cox 1987). 
 
\section{Muons and magnetism} 
Muons are ideally suited to studying problems in 
magnetism. 
Implanted muons in magnetically ordered materials precess 
in the internal magnetic field and directly yield signals 
proportional to that magnetic field.  In this respect the 
muon behaves as a microscopic magnetometer.  The very large 
magnetic moment of the muon  makes it very 
sensitive to extremely small magnetic fields (down to $\sim 10^{-5}$~T) 
and thus is very 
useful in studying small moment magnetism.  It is also 
valuable in studying materials where the magnetic order 
is random or of very short range.  Since muons stop 
uniformly throughout a sample, each signal appears 
in the experimental spectrum with a strength proportional 
to its volume fraction, and thus the technique is helpful 
in cases where samples may be multiphase or incompletely 
ordered.  Because no spatial information is obtained 
(in contrast to diffraction techniques) single crystal 
samples are not essential (though they can be useful 
in certain cases) and experiments can often provide 
information on the magnetic order of certain materials 
where conventional 
magnetic neutron diffraction cannot be simply performed. 
 
The most straightforward application of the technique 
is to ferromagnets and antiferromagnets in which cases 
the muon is used to follow the temperature dependence 
of the internal field and to extract critical exponents 
(Schenck and Gygax 1995). 
An example of this type of experiment has already been 
presented in Figure~\ref{npnn}. 
The internal field at the muon site is due to a 
sum of the applied field (if used), the dipolar and demagnetization 
fields (which can be calculated from the magnetization) 
and the hyperfine field induced by the applied field. 
To extract quantitative information from $\mu$SR experiments 
it is necessary to know the muon-site and this can in 
some cases hinder the search for a straightforward interpretation 
of the data.  Usually there are a small set of possible 
interstitial sites which the muon can occupy and 
in favourable circumstances only one will be consistent 
with the observed data. 
Nevertheless the technique has been widely applied to 
magnetic materials (Schenck and Gygax 1995, Dalmas de R\'eotier 
and Yaouanc 1997) and has found great applicability 
to the study of heavy fermion systems (Amato 1997). 
These latter materials are based on rare-earth 
and actinide elements and show strong electronic 
correlations between localized $f$ moments and conduction 
electrons. They exhibit a subtle competition 
between the Kondo effect (by which magnetic moments 
are `mopped up' by the screening spin polarization of 
conduction electrons) and the magnetic Ruderman-Kittel-Kasuya-Yosida 
(RKKY) interactions (which couple magnetic moments through 
the conduction electrons).  Depending on the relative 
strength of these effects, magnetic order can be observed 
at low temperature but it is often incommensurate or random 
and can be associated with 
very small static moments.  The sensitivity of the muon 
to tiny fields and the absence of a muon quadrupolar moment 
(which can complicate analogous NMR experiments) has 
led to muons being extensively utilised in this field (Amato 1997).

\begin{figure} 
\vspace{2.6cm} 
\includegraphics{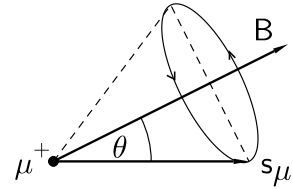} 
\caption{ 
Muon-spin precession with a magnetic field $B$ applied 
at an angle $\theta$. 
} 
\label{spinprec} 
\end{figure} 
 
\begin{figure} 
\vspace{16.4cm} 
\includegraphics{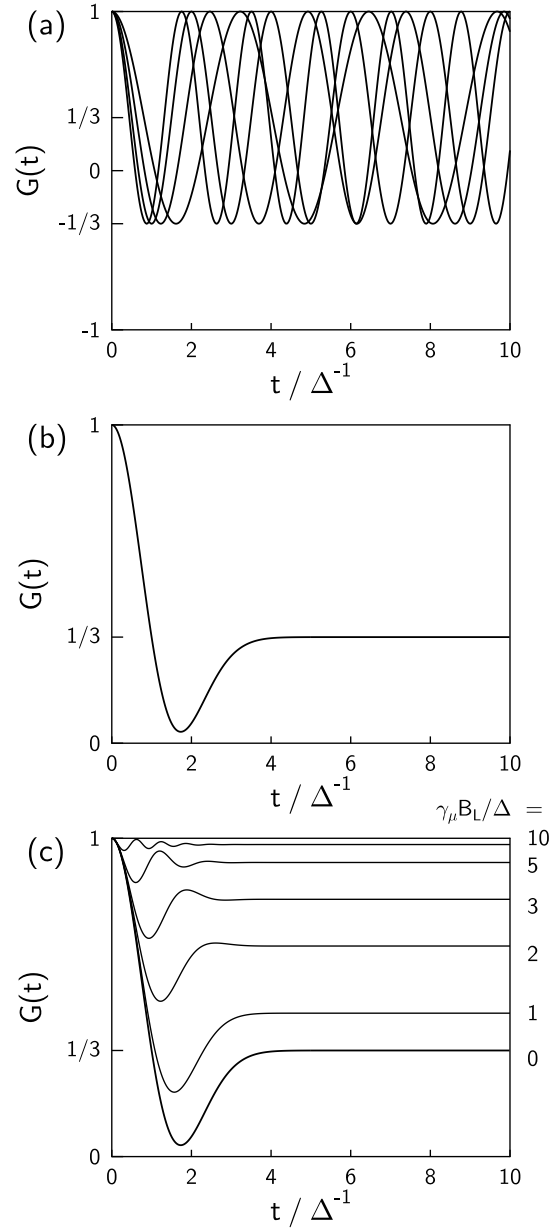} 
\caption{ 
(a) The time evolution of the muon-spin polarization 
for equation~\ref{sp1} with different values of the 
magnitude of the local field $\vert B \vert$. 
(b) The averaging of terms from (a) yields the 
Kubo-Toyabe relaxation function (equation~\ref{sp2}) 
with its characteristic dip and recovery to a value of 
$1/3$. 
(c) The effect of a longitudinal magnetic field $B_L$. 
The time is measured in units of $\Delta^{-1}$ and 
the longitudinal field values shown are in units of 
$\Delta/\gamma_\mu$. 
} 
\label{kubo} 
\end{figure}

To understand the ability of the muon to study randomness 
and dynamics in magnetism it is helpful to consider further 
some aspects of spin precession. 
If the local magnetic field at the muon-site is at an angle of 
$\theta$ to the initial muon spin-direction at the moment 
of implantation, the 
muon-spin will subsequently precess around the end of a cone of semi-angle 
$\theta$ about the magnetic field (Figure~\ref{spinprec}). 
The normalized decay positron asymmetry will be given by 
\begin{equation} 
 G(t)=\cos^2\theta + \sin^2\theta\cos(\gamma_\mu Bt) . 
\end{equation} 
If the direction of the local magnetic field is entirely 
random then averaging over all directions would yield 
\begin{equation} 
G(t)={1 \over 3} + {2 \over 3}\cos(\gamma_\mu Bt) . 
\label{sp1} 
\end{equation} 
If the strength of the local magnetic field is taken from 
a Gaussian distribution of width $\Delta/\gamma_\mu$ centred around 
zero, then a straightforward averaging over this distribution 
gives 
\begin{equation} 
G(t)={1 \over 3} + {2 \over 3}e^{-\Delta^2t^2/2}(1-\Delta^2t^2) , 
\label{sp2} 
\end{equation} 
a result which was first obtained by Kubo and Toyabe 
(Kubo and Toyabe 1967) as an entirely theoretical exercise. 
This relaxation function is illustrated in Figure~\ref{kubo}(b). 
Its origin is indicated schematically in Figure~\ref{kubo}(a) 
which shows a number of curves of equation~\ref{sp2} 
for different values of the internal field $B$.  Initially 
they all do roughly the same thing (i.e.\ fall from 1 to a minimum value 
and then increase) but after a short time they dephase with respect 
to each other.  Hence their average, the Kubo and Toyabe 
relaxation function, would be expected to fall from unity 
to a minimum and then recover to an average value, in this 
case to one-third.

This relaxation function is often observed experimentally, 
for example in experiments on copper at 
50~K with zero applied field (Luke \ea 1991). 
At this temperature the muon is stationary 
and precesses in the field due to the neighbouring 
nuclear dipoles which are randomly orientated with 
respect to each other and give rise to a field distribution 
each component of which is Gaussian distributed about zero. 
(At higher temperatures this is not seen because of thermally 
activated muon hopping while at lower temperatures an effect 
known as quantum diffusion may occur, see section~\ref{dynamics}.)

If the form of the distribution of internal fields 
was different then this would affect the form of the 
observed muon-spin time evolution.  For example, in 
a material with a spin-density wave which is incommensurate 
with the crystal lattice, there will be 
a sinusoidal modulation of the internal field 
which the muons will randomly sample.  In this case 
one finds that the muon-spin relaxation follows a Bessel 
function (Major \ea 1986, Amato 1997).  Such an effect has been observed in 
chromium (Major \ea 1986) and also in 
an organic system, (TMTSF)$_2$PF$_6$, which exhibits 
a spin-density wave ground state (Le \ea 1993). 
 
If there is an almost uniform static internal field in the sample, 
but there is a slight variation from site to site, 
different muons will precess at slightly different 
frequencies and become progressively dephased so that 
the oscillations in the data will be damped. 
If the field varies a great deal the damping could be so 
large that no oscillations can be observed. 
However this effect could also be caused by 
fluctuations either of the internal field, because of some 
intrinsic property of the sample, of 
the muon's position, or because of muon diffusion. 
One method of distinguishing between these possibilities 
is to apply a magnetic field 
in the longitudinal direction, parallel to the 
initial muon-spin direction. 
For example, this modifies the 
Kubo-Toyabe relaxation function as shown in 
Figure~\ref{kubo}(c), causing the ``${1 \over 3}$-tail'' 
to increase since in this case the muons precess in 
both the internal field and the applied field. 
Since the magnetic field $B_L$ is applied along the initial muon-spin 
direction then in the limit of large $B_L$ the muon-spin is held 
constant and does not relax from a value near unity. 
This technique is useful to distinguish the effects 
of inhomogeneous line broadening (a distribution of 
static internal fields as considered above) and fluctuations 
because the two have very different behaviours in 
a longitudinal field (Hayano \ea 1979).

\begin{figure} 
\vspace{5.6cm} 
\includegraphics{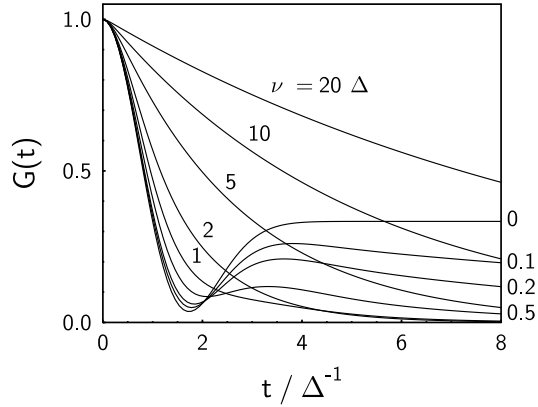} 
\caption{ 
The relaxation function for a muon hopping at rate $\nu$. 
After each hop the value of the internal field is taken 
from a Gaussian distribution about zero with width $\Delta/\gamma_\mu$. 
The curve for $\nu=0$ corresponds to the zero-field 
Kubo-Toyabe relaxation function of equation~\ref{sp2}. 
} 
\label{dkt} 
\end{figure}

The effects of muon hopping on the relaxation are shown in 
Figure~\ref{dkt}.  The different traces are for different 
hopping rates $\nu$.  When $\nu=0$ we recover the 
zero-field Kubo-Toyabe curve of equation~\ref{sp2}. 
For fast hopping the relaxation of the muon-spin becomes 
dominated by the hopping process and the relaxation is exponential. 
The relaxation rate goes down as the dynamics get faster.  This 
effect is known as motional narrowing because it is essentially 
identical to the effect of the same name in NMR spectroscopy. 
Motional effects narrow NMR linewidths which are measured 
in the frequency domain; consequently in the time domain 
(in which data from $\mu$SR experiments are usually considered) 
this corresponds to a motional widening and a reduction in 
relaxation. 
 
For slow hopping, very little effect is observed at very short times 
times but a large sensitivity to weak hopping is 
observed in the ${1\over 3}$-tail which is observed at 
longer times.  This sensitivity 
to slow dynamics via the behaviour of the tail 
of the relaxation function observed at long times 
permits a measurement of dynamics over a 
very large range of time scale. 
 
If a longitudinal magnetic field is applied then it has 
a large effect on the muon relaxation if the dynamics 
are weak but much less of an effect if the dynamics are 
fast.  Thus by a careful combination of zero-field 
and longitudinal-field experiments the nature of the 
internal field distribution can be extracted. 
 
Such an analysis has been done for many systems 
in which the issue of the presence of dynamics and/or 
magnetic order has profound consequences.  A good 
example is that of 
the spin ladder cuprates (Kojima \ea 1995). 
Sr$_2$Cu$_4$O$_{6}$ has a crystal structure such that 
Cu$^{2+}$ spins are arranged in weakly coupled 
2-leg ladders. $\mu$SR experiments show 
no magnetic order down to very low temperatures. 
Sr$_4$Cu$_6$O$_{10}$ 
is a 3-leg ladder and $\mu$SR finds static magnetic order 
appearing at $\sim$52~K.  This 
is 
consistent with theoretical predictions of a 
non-magnetic ground state in even-leg-number systems 
(due to the formation of spin-singlets between pairs 
of spins on each rung) but a magnetic ground state 
in odd-leg-number systems (Kojima \ea 1995). 
Similar types of experiments have been performed on SrCr$_8$Ga$_4$O$_{19}$, 
a material
which has magnetic moments lying in an arrangement known as a 
Kagom\'e lattice which is highly frustrated in the
sense that there is no unique ground state.  These measurements
have demonstrated the presence of dynamic spin fluctuations 
which persist down to at least 0.1~K (Uemura \ea 1994). 
These strong dynamics which remain at very 
low temperatures  are 
characteristic of a spin-liquid state in 
which a small number of unpaired spins 
migrate in a sea of singlet pairs. 
Some compounds with one-dimensional chains of antiferromagnetically 
coupled spins 
can show an effect known as the spin-Peierls transition. 
Below a transition temperature the chain dimerises, 
pairing alternate spins into singlets and opening up 
a gap in the excitation spectrum, drastically 
changing the dynamics.  $\mu$SR studies have 
been performed on both inorganic and organic 
systems which show this effect (Lappas \ea 1994, Blundell \ea 1997).

A celebrated example of a frustrated spin system is a 
spin glass, such as the dilute alloy spin glass prepared 
by dissolving small concentrations of Mn in a Cu matrix. 
In these system dilute magnetic impurities couple via 
an RKKY exchange interaction which alternates in sign 
as a function of distance.  Because the magnetic impurities 
are present at random, these materials cannot show long 
range order but have built-in frustration.  When cooled 
one observes a slowing down of spin fluctuations 
and a divergence in the correlation time between 
Mn spin fluctuations at the spin glass temperature. 
Below this temperature, a static component of the 
local field is observed with muons, corresponding to 
some degree of spin freezing, with each Mn spin having its 
own preferred orientation, although fluctuating around this (Uemura \ea 1985). 
Above the spin glass temperature muon-spin relaxation measurements 
have been used 
to follow the spin glass dynamics and to directly 
extract the form of the autocorrelation 
function of the spins (Campbell \ea 1994, Keren \ea 1996). 
 
An interesting situation occurs when a muon hops in an 
antiferromagnetically 
ordered lattice.  With no hopping, and if there is one muon site 
close to each spin in the system, there will be a single 
precession frequency. 
Because the sign of the internal field 
reverses every time the muon hops, the muon will precess one way and 
then, after hopping, precess back again.  As the hop rate increases 
the oscillations are progressively destroyed until eventually 
no relaxation is possible (Keren \ea 1993).  This effect has been 
observed experimentally in the cuprate compound 
Ca$_{0.86}$Sr$_{0.14}$CuO$_2$ (Keren \ea 1993).

\section{Muons and superconductivity} 
One of the most fruitful areas of recent research 
with muons has been in the area of superconductivity. 
The last couple of decades have witnessed 
a renaissance in this field following 
the discovery of high-temperature superconductors, 
organic superconductors, borocarbide superconductors, 
and even superconductors based on C$_{60}$ buckyballs. 
To understand the usefulness of muons, recall that 
the two important lengthscales in superconductors are the 
penetration depth, $\lambda$, which controls the 
ability of the superconductor to screen magnetic fields, 
and the coherence length, $\xi$, which controls the lengthscale 
over which the order parameter can vary without undue energy 
cost.  If the former is sufficiently greater than the latter 
(the condition is that $\lambda > \xi/\sqrt{2}$) the material is 
a type II superconductor which if cooled through 
its transition temperature, $T_{\rm c}$, in 
an applied magnetic field remains superconducting everywhere 
except in the cores of the superconducting vortices which
usually are arranged in a triangular lattice. 
Each vortex is associated with a magnetic flux equal to 
one flux quantum $\Phi_0=h/2e$.  The distance between vortices, $d$, 
is such that the number of vortices per unit area $2/\sqrt{3}d^2$ equals 
the number of flux quanta per unit area $B/\Phi_0$.  Thus $d \propto B^{-1/2}$. 
In general the vortex lattice will be incommensurate with the crystal 
lattice and, except at very high magnetic field, the vortex cores 
will be separated by a much larger distance than the unit-cell dimensions. 
Implanted muons will sit at certain crystallographic sites 
and thus will randomly sample the field distribution of the vortex lattice. 
 
\begin{figure} 
\vspace{7cm} 
\includegraphics{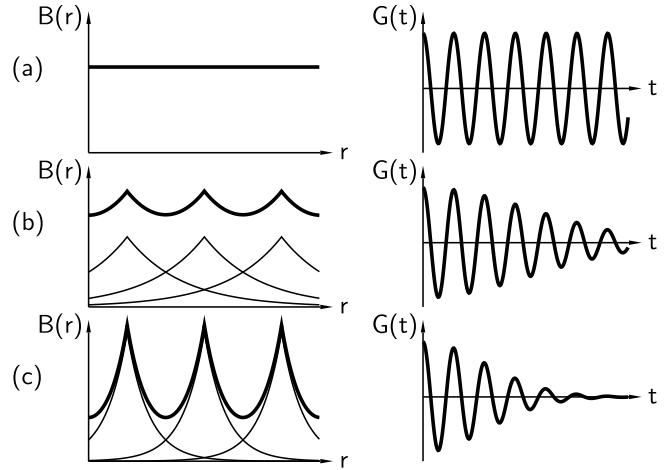} 
\caption{ 
The field distribution inside a superconductor 
as a function of position and the corresponding 
muon-spin relaxation function for three cases: 
(a) the normal state, (b) the superconducting 
state, (c) as (b) but with a shorter 
penetration depth. 
} 
\label{soup} 
\end{figure} 
 
In the normal state ($T>T_{\rm c}$) with a transverse field $B$, all muons 
precess with frequency $\omega=\gamma_\mu B$ (Figure~\ref{soup}(a)).  In the 
superconducting state however the muons implanted close to the vortex 
cores experience a larger magnetic field than those implanted 
between vortices.  Consequently there is a spread in precession 
frequency, resulting in a progressive dephasing of the observed 
precession signal (Figure~\ref{soup}(b)).  The larger the penetration 
depth, the smaller the magnetic field variation and the less pronounced 
the dephasing (compare Figure~\ref{soup}(b) and (c)). 
It turns out that this idea can be quantified (see Aegerter and Lee 1997) 
and that 
the relaxation rate $\sigma$ of the observed 
precession signal is related to the penetration depth 
using 
\begin{equation} 
\sigma  =  \gamma_\mu \langle B({\bf r}) - \langle B({\bf r}) 
\rangle_{\bf r}^{\vphantom{{1/2}}} 
\rangle_{\bf r}^{1/2} 
      \approx  0.0609 \gamma_\mu \Phi_0 / \lambda^2, 
\end{equation} 
where $B({\bf r})$ is the field at position ${\bf r}$ and 
the averages are taken over all positions. 
Thus the relaxation rate of the 
observed 
precession signal can be used to directly obtain the magnetic penetration depth. 
An advantage is that data are obtained from the bulk of the superconductor, 
in contrast to techniques 
involving microwaves which are only sensitive to effects at the 
surface. 
 
This principle has been applied to many different superconductors to 
extract both the penetration depth and its temperature dependence. 
This latter quantity is of great interest because it is 
a measure of the temperature dependence of the order 
parameter and can yield information concerning 
the symmetry of superconducting gap and hence the 
symmetry of the pairing mechanism. 
For example 
this approach has revealed unconventional pairing in a sample of 
the high temperature superconductor YBa$_2$Cu$_3$O$_{6.95}$ 
(Sonier \ea 1994). It is also possible to extract the 
vortex-core radius from a detailed analysis of the data 
(Yaouanc \ea 1997, Sonier \ea 1997). 
 
\begin{figure} 
\vspace{5.7cm} 
\includegraphics{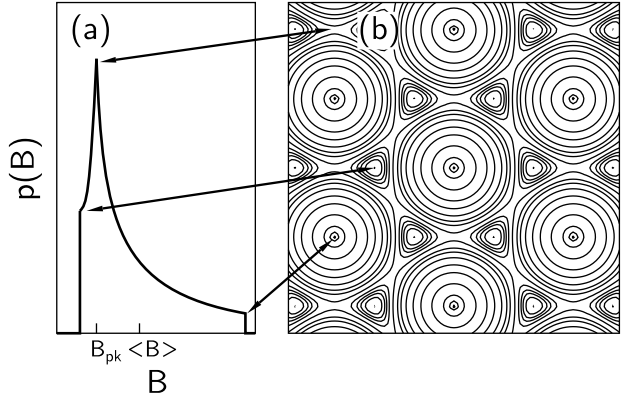} 
\caption{ 
(a) The field distribution $p(B)$ in the vortex lattice 
(contours of $B$ shown in (b)). 
} 
\label{fluxxy} 
\end{figure} 
 
A conventional type II superconductor exhibits 3 well-defined 
phases for $T<T_{\rm c}$: (1) a Meissner phase for $B<B_{c1}$, 
(2) a mixed or Shubnikov phase for $B_{c1}<B<B_{c2}$ (in which the 
magnetic field enters the superconductor in the form of 
well defined flux lines 
or vortices arranged in a lattice) 
and (3) the normal metallic phase for $B>B_{c2}$. 
In highly anisotropic systems the vortex lattice is no 
longer a system of rigid rods but should be considered 
as a system of flexible interacting lines.  A useful picture 
is that of a 
weakly coupled stack of quasi-two-dimensional (q2D) 
``pancake'' vortices, each one confined to a superconducting 
plane (Blatter \ea 1995, Clem 1991). 
The phase diagram is thus substantially altered to take account 
of field and temperature dependent changes in the vortex 
lattice itself.  At low $T$ and low $B$ the stacks resemble 
conventional vortex lines. 
Above a characteristic temperature $T_b$, but still below 
that at which superconductivity is destroyed, 
the vortex lattice is broken up by thermal fluctuations 
(Clem 1991) 
(this is called vortex lattice 
melting). 
At low $T$, but this time increasing $B$, 
the energetic cost of interlayer deformations of the lattice 
(local tilting of the lines) 
is progressively outweighed by the cost of intralayer deformations within the 
superconducting plane (shearing). 
Above a crossover field $B_{\rm cr}$ the vortex lattice enters a more 
two-dimensional regime. 
Thus in anisotropic systems we may 
expect temperature and field dependent transitions in which the 
vortex lattice is destroyed. 
When muons are implanted into 
a superconductor in a field $B_{\rm applied}$ one can directly measure 
the field distribution $p(B)$ which is given by 
$p(B)=\langle \delta (B - B({\bf r})) \rangle_{\bf r}$ and is the 
probability that a randomly chosen point in the sample has field $B'$ 
(Brandt 1988, Aegerter and Lee 1997). 
This is shown in Fig.~\ref{fluxxy}(a) for an ideal vortex line lattice. 
The distribution 
is highly asymmetric, the high field ``tail'' corresponding to regions 
of the lattice close to the vortex 
cores (see Fig.~\ref{fluxxy}(b)). 
The 
maximum 
of the distribution occurs at $B_{\rm pk}$, which lies  below the mean field 
$\langle B \rangle$ (see Fig.~\ref{fluxxy}(a)). 
Such lineshapes have been observed at low temperatures and fields 
in various anisotropic 
superconductors using $\mu$SR, including the high 
temperature superconductors (Lee \ea 1993, Aegerter and Lee 1997) 
and also in organic superconductors (Lee \ea 1997). 
In both case it is found that the vortex lattice can be melted 
with temperature at $T_{\rm b}$ or can cross into a two-dimensional 
regime at fields above $B_{\rm cr}$.  Both transitions can be 
followed by measuring the field and temperature dependence of 
the $p(B)$ line shapes. 
 
\begin{figure} 
\vspace{7.3cm} 
\includegraphics{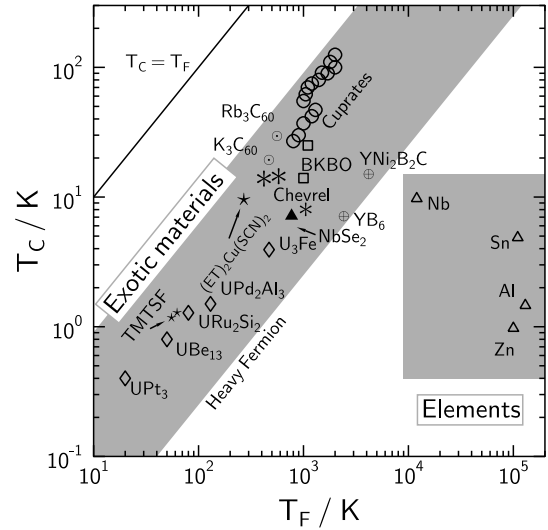} 
\caption{ 
The Uemura plot, showing the correlation between 
the superconducting transition temperature $T_{\rm c}$ 
and the Fermi temperature $T_{\rm F}$.  The exotic 
superconductors fall within a common band. 
Adapted from Uemura \ea 1991 and Hillier and Cywinski 1997. 
} 
\label{uemura} 
\end{figure} 
 
The London formula for the zero temperature 
limit of the penetration depth $\lambda(0)$ 
yields in the clean limit (the mean free path much 
bigger than the coherence length) the relation: 
$\lambda(0) \propto \sqrt{m^*/n_s(0)}$ where 
$m^*$ is the effective mass and $n_s(0)$ is the 
density of superconducting electrons. 
This can be combined with measurements of the Sommerfeld 
constant to yield a value for the Fermi temperature $T_{\rm F}$. 
Thus from muon measurements it is possible to 
plot a diagram showing the relationship between the 
Fermi and critical temperatures for a range of 
superconductors (Uemura \ea 1991).   This picture (see Figure~\ref{uemura}) 
has come to be called a Uemura plot and shows a clear correlation 
between $T_{\rm c}$ and $T_{\rm F}$ for the heavy fermion, organic, fullerene 
and Chevrel phase superconductors.  The conventional, elemental 
superconductors lie away from this correlation and have values 
of $T_{\rm c}/T_{\rm F}\sim 10^{-3}$ (for the ``exotic'' superconductors 
this value is one or two orders of magnitude larger). 
This correlation has been interpreted as evidence that the 
exotic superconductors may be close to Bose-Einstein condensation 
which is expected to occur at a temperature $\sim T_{\rm F}$ 
(Uemura \ea 1991).  Whether or not this speculation is correct, 
it is expected that this remarkable correlation should constrain 
theories to explain the superconductivity in these various exotic 
systems and is possibly suggestive of a common mechanism lying behind 
them.

\section{Muons as active probes} 
\label{dynamics} 
In almost everything we have considered so far we have been 
tacitly assuming that the muon is a passive probe and does 
not disturb its surroundings.  If it has been sensitive 
to dynamics, we have believed that the muon takes no part 
in them itself.  This in fact is very often true.  However, 
there are a number of situations in which the muon 
plays an active r\^ole.  For example, at high temperature in copper 
the muon diffuses from interstitial site to interstitial site.  In this 
case the major component of the observed depolarization is due to 
the muon motion.  In semiconductors the muonium can undergo 
charge and spin exchange with conduction electrons and thereby one 
measures dynamics with which the muon itself is intimately involved. 
 
\begin{figure} 
\vspace{3.6cm} 
\includegraphics{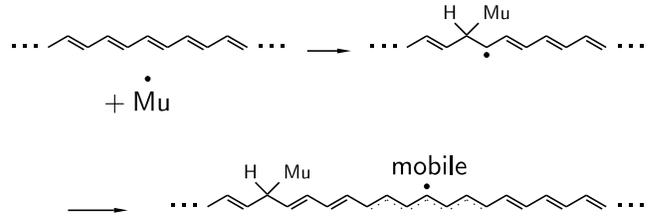} 
\caption{ 
Muonium interaction with trans-polyacetylene to produce 
a diamagnetic radical and a mobile neutral 
soliton. 
} 
\label{poly} 
\end{figure}

An extreme case where the muon plays a strongly 
active r\^ole is found in conducting polymers (Hayes 1995). 
Figure~\ref{poly} 
shows the reaction between muonium and trans-polyacetylene (Nagamine \ea 1984) 
which produces a diamagnetic, neutral muon defect and a highly 
mobile unpaired spin.  This soliton diffuses up and down 
the chain but cannot cross the muon defect which acts as a barrier. 
Every time the soliton briefly revisits the muon, the muon-electron 
hyperfine coupling is turned on and then off, so that successive 
visits progressively relax the muon polarization.  Measurement of the field 
dependence of this relaxation  yields 
the spectral density function associated with the defect random walk 
and can be used to infer the dimensionality 
of the soliton diffusion (Nagamine \ea 1984).  This 
occurs because the relaxation rate is connected with the noise 
power, $J(\omega_\mu)$, in the fluctuating magnetic field at the muon Larmor 
frequency, $\omega_\mu$, associated with that particular magnetic field. 
Sweeping the magnetic field allows one to extract the 
frequency distribution of the fluctuations. In other polymers, 
such as polyaniline, the muon-induced defect is a negatively charged 
polaron.  Muons are uniquely sensitive to the motion of this defect 
in undoped materials (Pratt \ea 1997) and in contrast to transport studies, 
which are 
inevitably dominated by the slowest component of the transport process, 
can provide information on the intrinsic transport processes governing 
the mobility of an electronic excitation along a chain.

In more conventional materials it is the motion of the 
muon itself which is of special interest. 
The dynamics of light atoms such as hydrogen and muonium, 
or particles such as the proton and muon, 
are worthwhile to study because they can provide a stringent 
test for theories on the quantum motion of defects 
and interstitials (Storchak and Prokof'ev 1998). 
The smaller mass of the muon 
leads to larger tunnelling matrix elements to neighbouring 
sites and thus enhances the quantum mechanical 
nature of the motion.  Furthermore, because muon-muon 
(or muonium-muonium) interactions can be neglected, 
the intrinsic nature of the dynamics can be followed 
without the complications that can be found in 
studying the corresponding proton or hydrogen case. 
 
\begin{figure} 
\vspace{3.5cm} 
\includegraphics{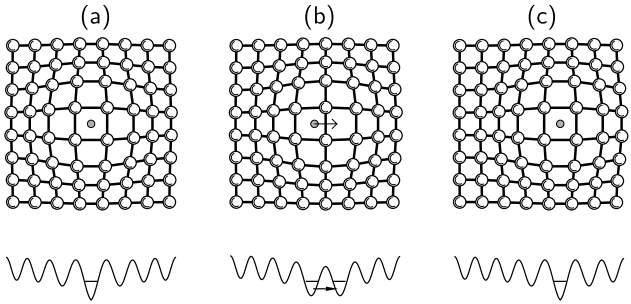} 
\caption{ 
Diffusion process of a muon by phonon-assisted tunnelling. 
(a) The muon is stable in an interstitial site and the local 
distortion leads to self-trapping so that its zero-point energy 
level lies a little lower than the neighbouring site. 
(b) Thermal fluctuations provide the opportunity for 
a coincidence configuration whereby tunnelling is allowed, 
leading to (c) a new stable configuration. 
(Adapted from Cox \ea 1987). 
} 
\label{lattice} 
\end{figure}

In inorganic materials the muon will usually come 
to rest at an interstitial 
site.  The stability of that site will depend on the 
depth of the potential well.  It is of interest to 
discover whether local diffusion is possible between 
interstitial sites.  Another process is trapping 
and release from deep potential wells associated with 
imperfections or defects.  The muon jump rates 
are found to be about ten times higher than the corresponding 
proton jump rates, consistent with the lighter muon mass. 
Hopping is, as expected, assisted by phonons and thus 
rises with temperature, following an approximately 
activated behaviour [the hop rates are proportional 
to $T^{-1/2}\exp(-E_a/k_BT)$ where $E_a$ is an activation 
energy (Flynn and Stoneham 1972)]. 
This occurs because the muon is initially `self-trapped' 
by its own local distortion of the lattice (Figure~\ref{lattice}) 
and a tunnelling transition is only possible if, by the 
thermal fluctuations of the lattice which occur because 
of phonons, two neighbouring energy levels coincide 
[the coincidence configuration, see Figure~\ref{lattice}(b)]. 
The muon can then tunnel through the barrier and becomes 
self-trapped in the next site. 
 
Lower temperature produces fewer phonons and hence the 
hop rate falls as the temperature is reduced. 
However as the temperature is lowered further, a peculiar effect is observed: 
the muon hop rate falls to a minimum and then begins to rise again. 
In the very low temperature regime the phonons appear to 
be hindering hopping rather than helping it, as they do 
at higher temperatures.  The reason is that 
at low temperatures coherent tunnelling is possible i.e.~the muon 
is in a band state.  Phonons now are responsible for inelastic 
scattering which destroys the coherence of this 
delocalized state.  This coherent effect 
is known as quantum diffusion (for a review, see Storchak and Prokof'ev 1998). 
 
\begin{figure} 
\vspace{9.5cm} 
\includegraphics{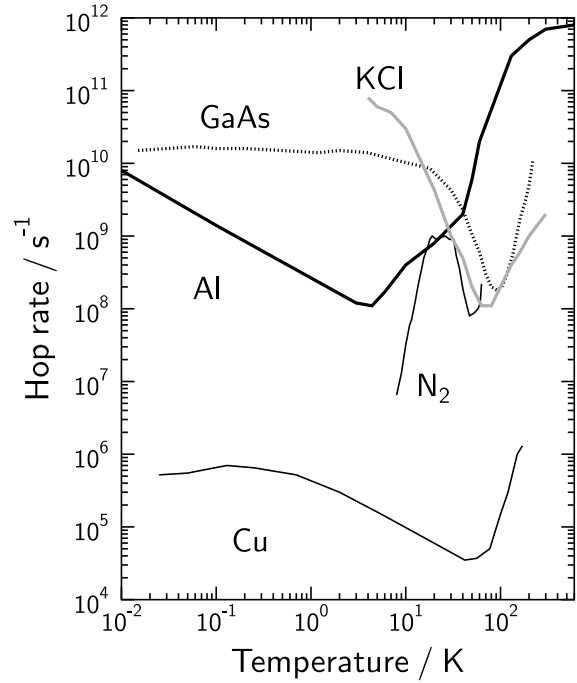} 
\caption{ 
Muon hop rates as a function of temperature for various materials. 
Copper (Luke \ea 1991), aluminium (Hartmann \ea 1988), 
gallium arsenide (Kadono \ea 1990), potassium chloride (Kiefl \ea 1989), 
and solid nitrogen (Storchak \ea 1994). 
} 
\label{hop} 
\end{figure} 
 
Experimental data for various materials are shown in Figure~\ref{hop} 
and although there are large differences in the size of 
the hop rate and the detailed form of the temperature dependence, 
all show an increasing hop rate at high temperature consistent with 
activated behaviour and a decreasing hop rate at low temperature 
consistent with quantum diffusion. 
An early theoretical treatment of this latter effect predicted 
that the low temperature hop rate would follow an inverse power 
law $T^{-\alpha}$, where the exponent $\alpha$ was large, 
typically $\sim 9$ (Kagan and Klinger 1974). 
Experiments on Cu and Al (see Figure~\ref{hop}) showed a more 
modest behaviour with $\alpha\sim$0.6--0.7. 
However it was shown that this could be explained 
by considering the effect of the conduction electrons 
in a metal, which screen the muon and cannot react fast 
enough to the diffusing particle and follow it adiabatically. 
This produces a net drag which reduces the particle hop 
rate and weakens the temperature dependence.  A detailed 
theory of this effect produces agreement with experiment 
(Kondo 1984, Yamada 1984, Kagan and Prokof'ev 1986). 
 
This dominant r\^ole of the electrons has been demonstrated by 
ingenious experiments on aluminium (Karlsson \ea 1995). 
In its superconducting state, 
the presence of the gap in the electronic spectrum 
effectively decouples the electron bath from the muon. 
At low temperature the superconductivity can be removed by 
applying a sufficiently large magnetic field 
to the sample.  This dramatically reduces the muon diffusion 
rate because the closing of the gap reconnects the muon diffusion 
process to the electron bath, introducing drag. 
 
In insulators there are no conduction electrons to worry about 
and in this case the hopping particle is a neutral muonium 
atom, not a charged muon. 
The coherent muon hop rate for KCl (Figure~\ref{hop}) rises much 
more rapidly with decreasing temperature than for metals, and 
fits to an exponent $\alpha=3.3$ (Kiefl \ea 1989). 
This still does not quite fit with the earlier theory 
(Kagan and Klinger 1974) but agrees with more sophisticated 
treatments (see Storchak and Prokof'ev 1998) which take into 
account the measured phonon spectrum in KCl measured using 
neutron scattering. 
 
In semiconductors muonium is also formed (see section~\ref{muonium}) 
and very similar temperature dependence is found (Figure~\ref{hop}, 
Kadono \ea 1990) with $\alpha\sim 3$ below 100~K but the 
hop rate saturates below $\sim$10~K due to the presence of disorder. 
(Below this temperature the coherent tunnelling is dominated by 
the disorder, rather than the phonons, and is therefore temperature 
independent.)  Data are also shown for solid nitrogen in Figure~\ref{hop}. 
This material shows a very sharp increase in hop-rate with decreasing 
temperature ($\alpha\sim 7$, Storchak \ea 1994); as the 
temperature is lowered further it saturates and 
thereafter begins 
to decrease.  This low temperature effect is thought to be due 
to orientational ordering of the N$_2$ molecules but is not yet understood 
in detail.

\section{Conclusions} 
Muons have found extremely wide application in condensed matter 
physics, allowing physicists and chemists to study 
a variety of problems concerning 
magnetism, superconductivity, chemical kinetics, diffusion, 
molecular dynamics, and semiconductor physics. 
The interested reader is referred to a number of excellent 
specialist reviews for further details concerning 
the technique and its application. 
(Schenck 1985, Cox 1987, Brewer 1994, Karlsson 1995, 
Dalmas de R\'eotier and Yaouanc 1997). 
Currently there is also active interest in attempting to 
utilise muons to catalyse fusion (Ponomarev 1990). 
 
The availability of muons at proton sources in various places in the 
world has led to a great deal of activity in this field. 
This is 
because even though the proton source is often designed 
for some other purpose (the production of neutrons for 
scattering experiments or the production of mesons for 
nuclear and particle physics), a muon facility can be 
``tagged on'' for little additional cost.  The consequent 
reduction in proton intensity for the main purpose 
of the proton source is 
only a few per cent.

One current limitation of muon techniques is that they require 
bulk samples because the incident muons are formed with 
energy 
4~MeV and penetrate a few hundred micrometres into any 
sample.  Even with the use of degraders, surface studies 
have so far not been possible.  Recently however there has 
been considerable progress in the development of so-called 
`slow muon' beams in which the energy of the muon beam 
is reduced down to $\sim$1--10~eV.  This can be 
achieved by either moderation in thin layers 
of rare gas solid 
(Morenzoni \ea 1994) or by resonant ionization 
of thermal muonium (produced from the surface of 
a hot tungsten foil placed in a pulsed proton beam) 
by a pulsed laser source (Nagamine \ea 1995). 
The efficiency of these processes is at present 
rather too low to allow routine experiments on thin films to take place. 
Nevertheless as moderator efficiencies improve and the 
necessary technology is developed, this promises to be a fruitful area of 
future research.

\small 
\section*{References} 
\begin{list}{}{%
 \setlength{\labelwidth}{0pt}%
 \setlength{\labelsep}{0pt}%
 \setlength{\leftmargin}{10pt}%
 \setlength{\itemindent}{-10pt}%
 \setlength{\itemsep}{-\parsep}} 
 
\item Aegerter, C. M., and Lee, S. L., 
1997, {\it Appl. Mag. Res.} {\bf 13}, 75. 
 
\item Amato, A., 
1997, {\it Rev.\ Mod.\ Phys.}, {\bf 69}, 1119. 
 
\item 
Anderson, C. D., and Neddermeyer, S. H., 
1936, {\it Phys. Rev.}, {\bf 50}, 263. 
 
\item Bailey, J., Borer, K., Combley, F., Drumm, H., 
Krienen, F., Lange, F., Picasso, E., von Ruden, W., 
Farley, F. J. M., Field, J. H., Flegel, W., and Hattersley, P. M., 
1977, {\it Nature}, {\bf 268}, 301.

\item Blatter, G., 
Feigel'man, M. V., 
Geshkenbein, V. B., Larkin, A. I., and  Vinokur, V. M., 1995, 
{\it Rev. Mod. Phys.}, {\bf 66}, 1125. 
 
\item Blundell, S. J., Pattenden, P. A., Pratt, F. L., Valladares, R. M., 
Sugano, T., and Hayes, W., 
1995, {\it Europhys. Lett.}, {\bf 31}, 573. 
 
\item Blundell, S. J., Pratt, F. L., Pattenden, P. A., Kurmoo, M., 
Chow, K. H., Jest\"adt, T., and Hayes, W., 
1997, {\it J. Phys.: Condens. Matter}, {\bf 9}, 119.

\item Brandt, E. H., 1988, {\it  Phys. Rev. B}, {\bf 37}, 2349. 
 
\item Brewer, J. H., 1994, 
Encyclopaedia of Applied Physics {\bf 11}, p23 (VCH, New York). 
 
\item Campbell, I. A., Amato, A., Gygax, F. N., Herlach, D., 
Schenck, A., Cywinski, R., and Kilcoyne, S. H., 
1994, {\it Phys. Rev. Lett.}, {\bf 72}, 1291. 
 
\item Caso., C., {\sl et al.}, 1998, The 1998 Review of Particle 
Physics, 
{\it Eur. Phys. J.}, {\bf C3}, 1. 
 
\item Chaikin, P. M., and Lubensky, T. C., 1995, 
{\it Principles of Condensed Matter Physics} 
Cambridge University Press (1995).  
 
\item Chappert, J., 1984, in {\it Muons and Pions in Materials Research} 
ed. Chappert, J. and Grynszpan, R. I., (Elsevier, 1984). 
 
\item 
Chow, K. H., Lichti, R. L., Kiefl, R. F., Dunsiger, S., 
Estle, T. L., Hitti, B., Kadono, R., MacFarlane, W. A., 
Schneider, J. W., Schumann, D., and Shelley, M., 
1994, {\it Phys. Rev. B}, {\bf 50}, 8918. 
 
\item 
Clem, J. R., 1991, {\it Phys. Rev. B}, {\bf 43}, 7837. 
 
\item 
Conversi, M., Pancini, E., and Piccioni, O., 
1945, {\it Phys. Rev.}, {\bf 68}, 232. 
 
\item 
Conversi, M., Pancini, E., and Piccioni, O., 
1947, {\it Phys. Rev.}, {\bf 71}, 209. 
 
\item 
Cox, S. F. J., 
1987, {\it J. Phys. C}, {\bf 20}, 3187. 
 
\item 
Dalmas de R\'eotier, P., and Yaouanc, A., 
1997, {\it J. Phys.: Condensed Matter}, {\bf 9}, 9113. 
 
\item 
Flynn, C. P., and Stoneham, A. M., 
1972, {\it Phys. Rev. B}, {\bf 1}, 3966. 
 
\item 
Garwin, R. L., Lederman, L. M., and Weinrich, M., 
1957, {\it Phys. Rev.}, {\bf 105}, 1415. 
 
\item 
Hartmann, O., Karlsson, E., W\"ackelg\aa rd, E., Wappling, R., 
Richter, D., Hempelmann, R., and Niinikoski, T. O., 
1988, {\it Phys. Rev. B}, {\bf 37}, 4425. 
 
\item 
Hayano, R. S., Uemura, Y. J., Imazato, J., Nishida, N., 
Yamazaki, T., and Kubo, R., 
1979, {\it Phys. Rev. B}, {\bf 20}, 850. 
 
\item 
Hayes, W., 
1995, {\it Phil. Trans. R. Soc. Lond. A}, {\bf 350}, 249. 
 
\item 
Hess, V. F., 
1912, {\it Phys. Z.}, {\bf 13}, 1084.

\item 
Hillier, A. D., and Cywinski, R., 
1997, {\it Appl. Mag. Res.} {\bf 13}, 95. 
 
\item 
Kadono, R., Kiefl, R. F., Brewer, J. H., Luke, G. M., 
Pfiz, T., Riseman, T. M., and Sternlieb, B. J., 
1990, {\it Hyp. Int.}, {\bf 64}, 635. 
  
\item 
Kagan, Yu., and Klinger, M. I., 
1974, {\it J. Phys. C}, {\bf 7}, 2791. 
 
\item 
Kagan, Yu., and Prokofe'v, N. V., 
1986, {\it JETP}, {\bf 63}, 1276.

\item 
Karlsson, E. B., 1995 
{\it Solid State Phenomena}, Oxford. 
 
\item 
Karlsson, E., W\"appling, R., Lidstr\"om, S. W., Hartmann, O., 
Kadono, R., Kiefl, R. F., Hempelmann, R., and Richter, D., 
1995, {\it Phys. Rev. B}, {\bf 52}, 6417. 
 
\item 
Keren, A., Le, L. P., Luke, G. M., Sternlieb, B. J., 
Wu, W. D., Uemura, Y. J., Tajima, S., and Uchida, S., 
1993, {\it Phys. Rev. B}, {\bf 48}, 12926. 
 
\item Keren, A., Mendels, P., Campbell, I. A., and Lord, J., 
1996, {\it Phys. Rev. Lett.}, {\bf 77}, 1386. 
 
\item 
Kiefl, R. F., Kadono, R., Brewer, J. H., Luke, G. M., 
Yen, H. K., Celio, M., and Ansaldo, E. J., 
1989, {\it Phys. Rev. Lett.}, {\bf 53}, 90. 
 
\item 
Kiefl, R. F., Schneider, J. W., MacFarlane, A., Chow, K., 
Duty, T. L., Estle, T. L., Hitti, B., Lichti, R. L., 
Ansaldo, E. J., Schwab, C., Percival, P. W., Wei, G., 
Wlodek, S., Kojima, K., Romanow, W. J., McCauley, J. P., 
Coustel, N., Fischer, J. E., and Smith, A. B., 
1992, {\it Phys. Rev. Lett.}, {\bf 68}, 2708.

\item 
Kojima, K., Keren, A., Luke, G. M., Nachumi, B., 
Wu, W. D., Uemura, Y. J., Azuma, M., and Takano, M., 
1995, {\it Phys. Rev. Lett.}, {\bf 74}, 2812.

\item 
Kondo, J., 
1984, {\it Physica B+C}, {\bf 126}, 377.

\item 
Kubo, R., and Toyabe, T., 1967, in 
{\it Magnetic Resonance and Relaxation}, ed. Blinc, R., 
p810, 
(North-Holland, Amsterdam). 
 
\item 
Lappas, A., Prassides, K., Amato, A.,  Feyerherm, R.,  Gygax, F. N., and 
Schenck, A., 1994, {\it Z. Phys. B}, {\bf 96}, 223. 
 
\item 
Lattes, C. M. G., Occhialini, G. P. S., and Powell, C. F., 
1947, {\it Nature}, {\bf 160}, 453, 486.

\item 
Le, L. P., Keren, A., Luke, G. M., Sternlieb, B. J., 
Wu, W. D., Uemura, Y. J., Brewer, J. H., Riseman, T. M., 
Uspani, R. V., Chiang, L. Y., Kang, W., Chaikin, P. M., 
Csiba, T., and Gr\"uner, G., 
1993, {\it Phys. Rev. B}, {\bf 48}, 7284. 
7284 (1993).

\item 
Lee, S. L., Zimmermann, P., Keller, H., Warden, M., 
Savi\'c, I. M., Schauwecker, R., Zech, D., Cubitt, R., 
Forgan, E. M., Kes, P. H., Li, T. W., Menovsky, A. A., and 
Tarnawski, Z., 
1993, {\it Phys. Rev. Lett.}, {\bf 71}, 3862.

\item 
Lee, S. L., Pratt, F. L., Blundell, S. J., Aegerter, C. M., 
Pattenden, P. A., Chow, K. H., Forgan, E. M., Sasaki, T., 
Hayes, W., and Keller, H., 
1997, {\it Phys. Rev. Lett.}, {\bf 79}, 1563. 
\item 
Luke, G. M., Brewer, J. H., Kreitzman, S. R., Noakes, D. R., 
Celio, M., Kadono, R., and Ansaldo, E. J., 
1991, {\it Phys. Rev. B}, {\bf 43}, 3284. 
 
\item 
MacFarlane, W. A., Kiefl, R. F., Dunsiger, S., Sonier, J. E., 
Chakhalian, J., Fischer, J. E., Yildirim, T., and Chow, K. H., 
1998, {\it Phys. Rev. B}, {\bf 58}, 1004.

\item 
Major, J., Mundy, J., Schmolz, M., Seeger, A., D\"oring, K. P., 
F\"urderer, K.,  Gladisch, M., Herlach, D., and Majer, G., 
1986, {\it Hyp. Int.}, {\bf 31}, 259. 
 
\item 
Morenzoni, E., Kottmann, F., Maden, D., Matthias, B., 
Meyberg, M., Prokscha, T., Wutzke, T., and Zimmermann, U., 
1994, {\it Phys. Rev. Lett.}, {\bf 72}, 2793. 
 
\item 
Nagamine, K., Ishida, K., Matsuzaki, T., Nishiyama, K., 
Kuno, Y., and Yamazaki, T., 
1984, {\it Phys. Rev. Lett.}, {\bf 53}, 1763. 
 
\item 
Nagamine, K., Miyake, Y., Shimomura, K., Birrer, P., 
Iwasaki, M., Strasser, P., and Kuga, T., 
1995, {\it Phys. Rev. Lett.}, {\bf 74}, 4811. 
 
\item 
Patterson, B. D., 
1988, {\it Rev.\ Mod.\ Phys.}, {\bf 60}, 69. 
 
\item 
Ponomarev, L. I., 
1990, {\it Contemp. Phys.}, {\bf 31}, 219. 
 
\item 
Pratt, F. L., 
1997, {\it Phil. Mag. Lett.}, {\bf 75}, 371. 
 
\item 
Pratt, F. L., Blundell, S. J., Hayes, W., Ishida, K., 
Nagamine, K., and Monkman, A. P., 
1997, {\it Phys. Rev. Lett.}, {\bf 79}, 2855. 
 
\item 
Reid, I. D., Azuma, T., and Roduner, E., 
1990, {\it Nature}, {\bf 345}, 328. 
 
\item 
Roduner, E., 
1993, {\it Chem. Soc. Rev.}, {\bf 22}, 337. 
 
\item 
Rossi, B., and Hall, D. B., 
1941, {\it Phys. Rev.}, {\bf 59}, 223. 
 
\item 
Schenck, A., and Gygax, F. N., 1995, 
in {\it Handbook of Magnetic 
Materials} {\bf 9} edited by Buschow, K. H. J., 
(Elsevier). 
 
\item 
Schenck, A., 1985, 
{\it Muon Spin Rotation: Principles and Applications in 
Solid State Physics} (Adam Hilger, Bristol). 
 
\item Sonier, J. E., Kiefl, R. F., Brewer, J. H., Bonn, D. A., 
Carolan, J. F., Chow, K. H., Dosanjh, P., Hardy, W. N., 
Liang, R., MacFarlane, W. A., Mendels, P., Morris, G. D., 
Riseman, T. M., and Schneider, J. W., 
1994, {\it Phys. Rev. Lett.}, {\bf 72}, 744. 
 
\item Sonier, J. E., Brewer, J. H., Kiefl, R. F., Bonn, D. A., 
Dunsiger, S. R., Hardy, W. N., Liang, R., MacFarlane, W. A., 
Miller, R. I., Riseman, T. M., Noakes, D. R., Stronach, C. E., and 
White Jr., M. F., 
1997, {\it Phys. Rev. Lett.}, {\bf 79}, 2875. 
 
\item 
Storchak, V. G., Brewer, J. H., and Morris, G. D., 
1994, {\it Phys. Lett. A}, {\bf 193}, 199. 
 
\item Storchak, V. G., and Prokof'ev, N. V., 
1998, {\it Rev.\ Mod.\ Phys.}, {\bf 70}, 929. 
 
\item 
Uemura, Y. J., Yamazaki, T., Harshman, D. R., Senba, M., and 
Ansaldo, E. J., 
1985, {\it Phys. Rev. B}, {\bf 31}, 546.

\item 
Uemura, Y. J., Le, L. P., Luke, G. M., Sternlieb, B. J., 
Wu, W. D., Brewer, J. H., Riseman, T. M., Seaman, C. L., 
Maple, M. B., Ishikawa, M., Hinks, D. G., Jorgensen, J. D., 
Saito, G., and Yamochi, H., 
1991, {\it Phys. Rev. Lett.}, {\bf 66}, 2665. 
 
\item 
Uemura, Y. J., Keren, A., Kojima, K.,  Le, L. P., Luke, G. M., 
Wu, W. D., Ajiro, Y., Asano, T., Kuriyama, Y., 
Mekata, M., Kikuchi, H., and Kakurai, K., 
1994, {\it Phys. Rev. Lett.}, {\bf 73}, 3306. 
 
\item 
Wulf, T., 
1910, {\it Phys. Z.}, {\bf 10}, 811.

\item 
Yamada, K., 
1984, {\it Prog.\ Theor.\ Phys.}, {\bf 72}, 195. 
 
\item 
Yaouanc, A., Dalmas de R\'eotier, P., and Brandt, E. H., 
1997, {\it Phys. Rev. B}, {\bf 55}, 11107. 
 
\item 
Xu, Q., and Brown, L. M., 
1987, {\it Amer. J. Phys.}, {\bf 55}, 23. 
 
\end{list} 
 
\end{document}